\documentclass{sig-alternate-05-2015}

\usepackage{color}
\usepackage{graphicx}
\usepackage{caption}
\usepackage{subcaption}

\usepackage[square,numbers]{natbib}

\let\OLDthebibliography\thebibliography
\renewcommand\thebibliography[1]{
  \OLDthebibliography{#1}
  \setlength{\parskip}{0pt}
  \setlength{\itemsep}{0pt plus 0.3ex}
}

\begin{document}

% Copyright
\setcopyright{acmcopyright}
%\setcopyright{acmlicensed}
%\setcopyright{rightsretained}
%\setcopyright{usgov}
%\setcopyright{usgovmixed}
%\setcopyright{cagov}
%\setcopyright{cagovmixed}

% DOI
%\doi{10.475/123_4}
%
%% ISBN
%\isbn{123-4567-24-567/08/06}

%Conference
%\conferenceinfo{ICSE '17}{May 20-28, 2017, Buenos Aires, Argentina}

%\acmPrice{\$15.00}

%
% --- Author Metadata here ---
%\conferenceinfo{ICSE}{'17 Buenos Aires, Argentina}
%\CopyrightYear{2007} % Allows default copyright year (20XX) to be over-ridden - IF NEED BE.
%\crdata{0-12345-67-8/90/01}  % Allows default copyright data (0-89791-88-6/97/05) to be over-ridden - IF NEED BE.
% --- End of Author Metadata ---

\title{Database as a Service - Current Issues and Its Future}
%\subtitle{[Extended Abstract]
%\titlenote{A full version of this paper is available as
%\textit{Author's Guide to Preparing ACM SIG Proceedings Using
%\LaTeX$2_\epsilon$\ and BibTeX} at
% \numberofauthors{1} %  in this sample file, there are a *total*
% % of EIGHT authors. SIX appear on the 'first-page' (for formatting
% % reasons) and the remaining two appear in the \additionalauthors section.
% %
%
 \author{
 \alignauthor
 Xi Zheng$^{1}$\\
 \affaddr{$^{1}$Department of Computing, Macquarie University, Sydney, Australia\\
 \email{james.zheng@mq.edu.au}
 }}

\maketitle
\begin{abstract}
With the prevalence of applications in cloud, Database as a Service (DBaaS) becomes a promising method to provide cloud applications with reliable and flexible data storage services. It provides a number of interesting features to cloud developers, however, it suffers a few drawbacks: long learning curve and development cycle, lacking of in-depth support for NoSQL, lacking of flexible configuration for security and privacy, and high cost models. In this paper, we investigate these issues among current DBaaS providers and propose a novel {\em Trinity Model} that can significantly reduce the learning curves, improve the security and privacy, and accelerate database design and development. We further elaborate our ongoing and future work on developing large real-world SaaS projects using this new DBaaS model.

\end{abstract}




%
%  Use this command to print the description
%

% We no longer use \terms command


\section{Introduction}
In cloud computing, Database-as-a-Service (DBaaS) is a service model adopted increasingly by organizations recently. Enterprise customers consider DBaaS as a significant initiative that will empower developing an application efficiently~\cite{lehner2010database,zheng2017big}. DBaaS moves database management system (DBMS) and data storage from a client-server design where the owner of data oversees DBMS and reacts to the queries, to a third-party cloud architecture where data administration is not controlled by the data owner. The market of the DBaaS is expanding significantly and the trend of adoption is not restricted to large scale organizations but also happens to small enterprises~\cite{lehner2010database,bienko2015ibm}.

DBaaS offers a few important features.
% It has no need to install, configure, store and maintain any local database servers. 
Firstly, It offers database management system as on-demand service for handling and controlling data. It also provides a ubiquitous access to the data over the network. Thus, consumers can access  abstract resources and accomplish preferred operations at any time at any where (so long having the internet connection). Another key feature provided by DBaaS is the scalability of underlying data so that service providers can manage fluctuations in the workload~\cite{gelogo2012database}. Furthermore, DBaaS providers RDBMS and NoSQL, each of which have its own specific features which are based on ACID (Atomicity, Consistency, Isolation, and Durability) and BASE (Basically Available, Soft state, Eventually consistency) properties respectively~\cite{stonebraker2010sql}. These features make it a better solution in a dynamic cloud environment. 

However, DBaaS also has some issues apart from above features.  The issues lie in data confidentiality, data integrity, data authenticity, data privacy, data trust, high cost model and lacking of support to hybrid models. In hybrid models, all or part of the application data are required to store in customer's premise~\cite{ferrari2009database,weis2011securing}. When it comes to concurrent access to the records in a distributed database for a large-scale cloud application, we face issues like data-locking, reads and writes conflicting~\cite{bernstein1981concurrency}.  
The concurrent issues can't be soved by the current DBaaS models as RDBMS lacks the demands of availability, scalability, quick data backup and data recovery, while NoSQL lacks consistency and efficiency for handling complex queries. 

A tight integration of RDBMS with NoSQL shall provide a good solution for the large-scale cloud applications' database needs and make the cloud applications more scalable. However, they are not supported by the current DBaaS providers~\cite{zhang2011middleware}. Besides lack of support for the integration of RDBMS with NoSQL, DBaaS also lacks support to remove the huge learning curve and resolve limited choice to use NoSQL tools offered by DBaaS providers. 
%JSON-like field-value pair documents can be taken as one of the instances with respect to the learning curve of NoSQL. 
With a diminished emphasis on consistency of data, No-SQLs can offer greater flexibility~\cite{banker2011mongodb}. NoSQLs tend to be more elastic, with regards to scale out and scale up, than relational systems. 
% An elastic database architecture enables business to rapidly scale up their database to hundreds of servers. NoSQLs generally support dynamic workloads and are amenable to distributed analytics because of the distributed architecture of the data layer. 
For instance, key-value store NoSQL represents an associative array of keys, mapping to a value. The objects stored within them are transparent to the user interacting with the database. Objects can only be referenced by their keys, and not directly referenced by their values. Key-value stores are stored internally as a hash map, and therefore are highly scalable. On the other hand, graph databases represent one of the least-used varieties of NoSQL databases~\cite{bienko2015ibm,gelogo2012database}. Graph databases excel at dealing with highly interconnected data (e.g., Facebook) and can trace the relationship between different data nodes. 
Similarly, spatial-temporal database supports spatial-temporal queries, which are suitable for the modeling of moving objects, development of constraint based formalisms and spatial-temporal models~\cite{chen2011incremental}.
Moreover, spatial-temporal databases are increasingly important with respect to data analytics and algorithms for clustering mining~\cite{djafri2002spatio}. Unfortunately, current DBaaS providers do not support this variety of NoSQL databases to cloud developers.

In this paper, we have the following contributions. This paper is the first study to investigate DBaaS issues (Section~\ref{sec:literatureReview}). We propose a novel solution, named {\em Trinity Model} as a more suitable and practical DBaaS model for building cloud applications especially for Software-as-a-Service (SaaS) providers (Section~\ref{sec:proposedSolution}). Finally Section~\ref{sec:conclusions} concludes the paper.
% We start by walking through a few key issues in the current DBaaS service models in Section~\ref{sec:literatureReview}. 
%Then it is followed by a survey with detailed results that focus on using different features provided by DBaaS in section 4 and then summarize the interview results from experts in section 5. 
% Then we propose the {\em Trinity} solution to overcome these challenges along with our ongoing and future work to develop a few large scale SaaS services using the new service model in Section~\ref{sec:proposedSolution}. %After that, we discuss about several threats to validity in section 7 and then 

%In the first paragraph talk about the popularity and promising future of using DaaS
%
%Second paragraph talk about some key features offered by DaaS
%
%Third paragraph talk about the challenges in DaaS mainly in turns of solving locking and write/read conflicts of RDBMS, integration with NoSQL, learning curve of writing efficient queries in RDBMS, learning curve of using NoSQL (especially graphical database for building social network applications and spatial/temporal database for building IoT applications).
%
%Fourth paragraph talk about our work - literature review, survey, interview, and propose new solutions to highlight our contribution. 
%
%Last paragraph to stress our study is the first to....... And also talk about the structure of the rest of the papers.

%
%\section{Methodology}
%\label{sec:methodology}
%\input{methodology}

\section{Literature Review}
\label{sec:literatureReview}
In this section,  due to brevity we walk through some key challenges of current DBaaS service models.The main challenges include long development cycle and learning curve, lack of support for tight integration of RDBMS and NoSQL,  high cost models, issues with security and privacy.

\subsection{Long Development and Learning Curve}
% In the current DBaaS models, variety of RDBMS are provided which includes SQL Server, Oracle, MySQL and many others. However, 
There are some challenges for RDBMS which are inherited and not solved by DBaaS models. The mass data generated dynamically in cloud computing applications requires the underlying relational database to record hundreds of millions of records in the storage in a table and the efficiency to run SQL query is extremely low~\cite{han2011novel}. 
% Users can scale a relational database by running it on more powerful and expensive nodes (servers). Nonetheless, 
% when scale-up becomes no longer economically and practically an option,  scale-out has to be used where the data must be distributed across multiple servers. This is when the complexity of relational databases starts to increase dramatically. 
In this regard, relational databases do not work easily in a distributed manner owing to the difficulties that are faced while joining their tables across the underlying distributed systems~\cite{leavitt2010will}. 
% Besides, with the relational databases, users must convert all data into tables. 
When the data doesn't fit easily into a table, the structure of a database can be complex, difficult and slow to work with~\cite{leavitt2010will}. Relational databases have been criticized for the strong typing of the relation schemas which have ultimately made difficult for altering the data model. Even minor changes to the data model have to be done carefully and this might also require downtime or reduced service levels~\cite{atzeni2013relational}. 

One of the key challenges when developing a multi-tenant application (e.g., SaaS service) is to maximize concurrent access~\cite{ferrari2009database}. When multiple transactions need to be executed concurrently in a database, there inevitably is a high chance of violating the Isolation and Consistency of the ACID properties. In order to ensure consistency and isolation within the database, the database system needs to control the interaction of operations between concurrent transactions.
%The most common concurrency issue is a lost update in which the modifications from one transaction is overridden by the modification of the next transaction~\cite{kyte2014locking}. The mechanisms of locking and multi-version concurrency control are incorporated to achieve concurrent controls. As such, locks are mechanisms used to regulate concurrent access to a shared resource when multi users are accessing the database and thus ensuring data integrity and data consistency by attaining read lock or write lock~\cite{kyte2014locking,bernstein1981concurrency}.  
% The cloud application developers have to understand these concurrency issues and plan scale-up and scale-out strategy very carefully, thus defeating the very purpose of using Database as a service where these level of database underlying complexity shall be hidden away from cloud application developers.
Current DBaaS models do not remove business from the need to hire database specific developers, architects, and database administrators , which are very costly for the business and significantly increases the development cycle and learning curve for the development team.

Similarly, NoSQL also faces issues and challenges regarding development overhead and learning complexity. 
For instance, learning Mongodb is not an easy task for novice developers for MongoDB, where an experienced developer has to go under specific training. 
% The official MongoDB University is the real example where the company has made training as an open-source to everyone who wants to learn, depending on different courses. 
Some most popular courses are ``MongoDB for developers" and ``Getting Started with MongoDB Cluster Management". 
% The duration for every course is minimum 7 weeks and an exam week~\cite{MUniversity}.
Moreover, since NoSQL databases do not work with SQL, they require manual query programming, which can be fast for simple tasks but it can be time-consuming for complex tasks~\cite{leavitt2010will}.
Relational databases natively support ACID properties while NoSQL databases do not support. As such, NoSQL databases do not natively provide the degree of reliability that ACID provides. Though not providing consistency enables better performance and scalability, however, it is a problem for certain types of applications and transactions~\cite{leavitt2010will}, using NoSQL alone creates non-trivial issues for cloud computing applications.

\subsection{Lack of support for tight integration of RDBMS and NoSQL}
% A good DBaaS must support database and work-loads of different sizes enhancing elastic scalability. The challenge arises when a database work-load exceeds the capacity of a single machine. In this regard, 
A DBaaS must support scale-out, where the responsibility of query processing and the query data are partitioned amongst multiple nodes to achieve higher throughput~\cite{curino2011relational}. 

NoSQL (Non-relational SQL) data stores today emerging as a new trend in contrast to relational databases. There are major features in NoSQL databases like management of non-transactional large streams of data, a quick access to key-value, MapReduce for big data analysis. Because of their inherited distributed architecture, NoSQL databases are highly scalable and available and also less rigid in terms of their data layout schema. This makes the read-write operations very simple and fast for NoSQL databases. 
In contrast to ACID properties, NoSQL data stores inherit CAP theorem from Brewer~\cite{pokorny2013nosql}:
i. Consistency: Every node that reads from database sees same data;
ii. Availability: Every request made should receive a response irrespective of success or failure;
iii. Partition Tolerance: Database can be accessible even after any node-failure in the network.
Due to these facts, NoSQL databases are able to serve large scale cloud application and multi-tenant services, and lead to minimum cost management and resource utilization improvement~\cite{pokorny2013nosql}.

%NoSQL databases are improving in acceptance for many applications particularly Big Data in analytics, high availability and scalability. NoSQL systems are particularly used to handle situations where applications not served as expected by RDBMS and often needs processing Big Data. NoSQL systems can be classified into graphical, document column and key-value pair. One another issue prominently seen is, there are no such standards APIs or query languages for accessing either NoSQL or RDBMS. These several NoSQL databases, each has its own form of API which causes hindrance in integration process with the RDBMS, such as SQL and JDBC. Integrating these systems with other reporting and enterprise applications needs additional development efforts. This reduces the system portability which is another issue in the integration process~\cite{lawrence2014integration}.  
Current DBaaS models neither support variety of NoSQL databases as required for the cloud developers, nor does the current models support a tight integration of RDBMS and NoSQL where ACID properties and CAP theorem can be utilized and integrated tightly to best suit the cloud developers specific needs.

\subsection{High Cost Model}
Cost is one of the prime factors on deciding which model to use  \cite{gelogo2012database}. Depending on budget and various purposes, business requires flexible cost models to deploy different applications. At the moment, the main DBaaS providers offers three major sub-models: machine resources, data transfer and data storage.
%he companies have different plans for different needs as per enterprise budget. There are many players in the market providing DBaaS solutions. Our research focuses on cost model of main dominating players namely Amazon's RDB (Relational Database), Google's BigTable, IBM's Cloudant and Microsoft's Azure SQL Database. Also, the emerging DBaaS players from china are leading their business on global level. The players Alibaba Cloud, Tencent Cloud and Baidu Cloud from china are also been compared here with their cost models. 
%The cost model comparison is made on 3 major sub costs: machine resources, data transfer and data storage. 
The payment method is in a way of ``Pay as you go". The users pay based on how much resources (CPU, Memory and bandwidth) they use.

The unit price for different cost models may seem low, however, it poses a non-trivial running cost for both cloud providers and business. We envision a new open-source DBaaS service middleware where the service model is not only more practical, but also more affordable. It is practical because DBaaS services are delivered based on open source middleware; it is affordable because data storage and data transfer costs are removed from payment. 
% business only need to pay the necessary machine usage and running costs (e.g., monthly rental fee which is constant).

\subsection{Security and Privacy Issues}

Another challenge lies in security and privacy~\cite{ferrari2009database,weis2011securing}. Securing data requires: data confidentiality to avoid disclosure of essential information; data integrity to protect data from the unauthorized modifications; data availability to recover data from failures~\cite{castano1995database}. Work in~\cite{munir2015security} presents a four-layer structure to ensure in-cloud data security: 1) user authentication; 2) accessing control over software and storage in the cloud; 3) efficient and reliable service to manage a database, which allows reuse of the queries residing in cache; 4) data storage layer where data is encrypted at storage and decrypted at retrieval.

To further improve the security and privacy in case critical data for all tenants are stored in centralized cloud resource, which is becoming a single point of failure, we propose some features for DBaaS. An ideal DBaaS model should allow sensitive data to be stored in clients' own premise and provide a tightly integrated application specific security (e.g., real time traffic classification~\cite{zhang2015robust}, secure communication~\cite{yusurvey},  and security operation center~\cite{valeur2004comprehensive,wen2010lightweight, wen2012cafs,zheng2016investigating}, more intuitive encryption~\cite{anvari2017generating},  runtime monitoring of abnomal behaviors~\cite{zheng2014braceassertion,zheng2014physically,zheng2015braceassertion,zheng2015perceptions,zheng2015verification,zheng2014state}, and differential privacy mechanism~\cite{dwork2006calibrating}), which we will elaborate more in Section~\ref{sec:proposedSolution}.

%We shall dig very deep into the challenges in using DaaS to develop cloud application in general.  
% highlighting the issues in cost models, development cycle,  and hybrid models especially regarding solving RDBMS lock issues, read/write conflicts (casued by Index), difficulty in integration with RDBMS with NoSQL, huge learning curve in MongoDB, graphical database, and spatial/temporal database.  But shall also cover other aspects, which I have highlighted in the previous meeting.  Please check the recorded audio which I have also pinned on the right panel in the general channel.
%
%Talk about what are the current solutions to tackle these challenges. 
%
%Formulate a list of hypothesis, which basically says cost models are high, development cycle too long, MongoDBd is very hard to use, and diffuclt to access graphical and spatial/spatial databases, hybrid models not supportec.  Plus other issues likesecurity, privacy..... These questions shall be re-iterated in the online survey and dig even deeper in the interview.

% From the literature review above, it is evident in current DBaaS models there are a few key shortcoming and research challenges associated, in the next section we propose Trinity architecture to deal with these challenges.

%\section{The On-Line Survey}
%\label{sec:survey}
%\input{onlineSurvey}
%
%\section{Interviews}
%\label{sec:interviews}
%\input{interview}

\section{Proposed Solution}
\label{sec:proposedSolution}
In this section, we describe {\em Trinity Model} and discuss our ongoing and future work in building large real-world SaaS projects using the model along with some key challenges.

\subsection{Architecture Design}
\begin{figure}[t]
\centering
\hspace{-4.7em}%
\begin{subfigure}[b]{.28\linewidth}
  \includegraphics[width=\linewidth]{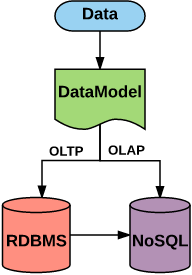}
   \caption{Architecture}
   \label{fig:trinity}
\end{subfigure}
\hspace{.3em}%
\begin{subfigure}[b]{.5\linewidth}  
  \includegraphics[width=1.4\linewidth, height=27mm]{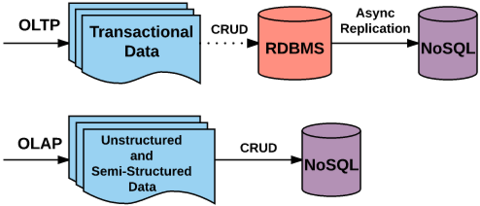}
   \caption{Processing}
   \label{fig:datahandling}
\end{subfigure}%
\caption{Architecture and Processing}
\label{fig:test}
\end{figure}

As shown in Figure.~\ref{fig:trinity}, the proposed model consists of three components: 1) Data Model; 2) RDBMS; 3) NoSQL.  
%
%
%
%figure \ref{fig:datahandling} 
%This proposed solution consists of following components:
%
%i.      Data Model
%
%ii.     RDBMS
%
%iii.    NoSQL

\begin{figure}[t]
\hspace{-1em}%
\includegraphics[width=0.5\textwidth, scale=0.4, height=68mm]{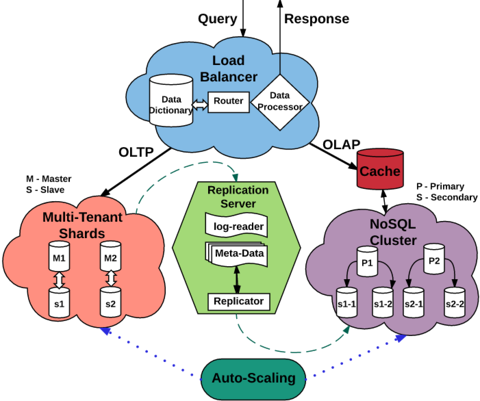}
\caption{Trinity Model Conceptual Design}
\label{fig:trinityconceptualdesign}
\end{figure}

The data model is used by information architects to define the type of data. The data can be either structured, semi-structured or unstructured. The model allows information architects to manipulate data that requires ACID properties and transactions. Thus, information architects can logically identify data schema, data relationships and data operations. They may read and write data without knowing the underlying technical details of RDBMS and NoSQL databases.
The data model specified by information architects allow {\em Trinity Model} to know how to handle the data with RDBMS and NoSQL databases.
As shown in Figure.~\ref{fig:datahandling},  transactions and structured data are automatically stored into RDBMS databases with a most efficient and scalable way, which is elaborated more in Figure.~\ref{fig:trinityconceptualdesign}. And these data will be replicated to NoSQL databases with very low latency for data retrieval purpose (mainly for data analytics).
For unstructured and semi-structured data, it will be stored in corresponding NoSQL databases (e.g., Document DB, Graphical DB, Spatial-Temporal DB) directly depending on the attributes of these data.
As shown in Figure.~\ref{fig:datahandling}, the separation of these data depending on their attributes allows our {\em Trinity Model} to handle Online Transaction Processing (OLTP) and Online Analytical Processing (OLAP).

% To be more specific, relational databases are designed for handling OLTP requests from cloud developers. The main reason for this is all the relational databases are ACID compliant, which plays a crucial part for OLTP. RDBMS handles all the structured data, the transactional data is processed and stored in relational database management systems such as MySQL, Oracle, etc. 
% On the other hand, NoSQL databases stores non-relational, i.e., semi-structured and no structured data. There are three advantages of doing so: NoSQLs have less rigid schema which is easy to host raw data for OLAP; NoSQLs choose data availability over consistency which makes it a good candidate for online queries; NoSQLs naturally embrace distributed architecture, so they have good scalability to deliver a high throughput.

%
%our {\em Trinity} model is able to store
%
%replicate this data from the relational model into a NoSQL model for reading. This results in a fast and secure large-scale data storage solution. As such, the relational database handles the transactional operations and stores the structured data which is replicated to NoSQL database and NoSQL handles semi-structured and unstructured data as shown in figure \ref{fig:datahandling}.

%\begin{figure}
%\centering
%
%\end{figure}
%\begin{figure}
%
%\end{figure}
%
%\begin{figure}
%\centering
%\includegraphics[width=30mm]{datasplit}
%\caption{Splitting OLTP and OLAP between RDBMS and NoSQL}
%\label{fig:datasplit}
%\end{figure}

%
%Before storing the actual data, one has to understand the data and observe the relationship between the other data sets. Different forms of data, either it is structured, semi-structured or unstructured, an application's data model shall handle all these forms for better scalability. From the figure \ref{fig:datasplit}, the data model of Trinity architecture resolves this for handling data processing in a better way by separating read/write requests and forwarding OLTP-to-RDBMS and OLAP-to-NoSQL. 

Figure \ref{fig:trinityconceptualdesign} is the architecture of our proposed  {\em Trinity Model}. 
It includes four major components: Load Balancer, RDBMS shards, Replication server, and NoSQL Cluster. The idea is to build a database with tight integration of RDBMS with NoSQL databases. We are ambitious to provide high availability and scalability to multiple tenants at the same time. We explain each components in details as follows.

The load balancer efficiently handles the request load on database servers. It maintains a data dictionary with meta-data that identifies database type (RDBMS or NoSQL), a router to identify the server location and forward the corresponding request, and a data processor that aggregates all the requested data from distributed systems before sending the response back to client. 

The RDBMS shards is to distribute the structured data that is identical for multiple tenants over various individual databases. This multi-tenant sharding helps maintain high volumes data, reduce overall workload on a single server. These sharded databases are then replicated as Master-Slave sets. Slaves act as backups to support failure over. An asynchronous replication is followed by default in the RDBMS. 
% The master node records all the write events into its binary log. Then this log events file will be used by the slave to update all the events into its own node to maintain consistency.

The replication server handles the data replication process from RDBMS to NoSQL. Replication of data from relational tables to schema-less collections is done by a live replication process based on changes in RDBMS logs. This is done in three stages: mapping schema to schema-less collections, extracting live changes of records from binary database logs, recorded logs are used to transfer the data.

NoSQL cluster handles all the OLAP requests. Each shard of RDBMS database can have a corresponding replica set in NoSQL cluster. Each replica set has multiple secondary nodes but only one primary node. These secondary nodes delivers high availability of data to the cloud tenants. In case a primary node goes offline, one secondary node is promoted to keep serving requests. 
% Even though all the members of replica set can be utilized for read operations, trinity model specifically assigns all write operations to primary and read operations to the secondary nodes to distribute workload.

To handle high concurrency, the {\em Trinity Model} caches recent queries on top of NoSQL. This improve the performance by avoiding repeatedly fetching data from NoSQLs. Auto-Scaling module provides current tenants with high availability of the service in the circumstances of network fluctuations and changes of workload. {\em Trinity Model} will respond to those needs and serve such requests by facilitating extra resources or by reconstructing the existing resources.

\subsection{Our Work and Research Challenges}
We are currently building the prototype of {\em Trinity Model} and helping one of the largest software powerhouses to develop Points of Sales (PoS), Customer Relationship Management (CRM), Supply Chain Management (SCM), and Recommendation and Data Analytics Systems (RDAS).  

Our first work is to conduct extensive survey and interview developers who use DBaaS across the world. The preliminary survey questions are available\footnote{Survey URL: https://www.surveymonkey.com/r/P7RPVVL}. Based on the survey results and participants' feedback,  we will conduct more in-depth interviews.  The results will help us thoroughly understand technical and research challenges in DBaaS models and validate some of our claims in this paper which we retrieved from the existing research literature.

Our second work is to develop the prototype and co-develop the PoS, CRM, SCM, and RDAS systems as SaaS services using the {\em Trinity Model}.  During this process, we will tackle the following research challenges:
\begin{list}{}
{
\leftmargin=1em
\itemsep=-.6em
\labelwidth=.5em
\topsep=-.4em  
}
\item[$\circ$] Create a practical and intuitive data model which can be used by information architects alone to define the data type, data relationship, and CRUD (Create, Retrieval, Update, and Deletion) operations on the data~\cite{schram2012mysql}. 
\item[$\circ$] Create a middleware which is able to replicate with very low latency data from RDBMS to NoSQL \cite{lin2013qos}.
\item[$\circ$] Incorporate Graphical database, Spatial Temporal database, and Document Databases into NoSQL model \cite{chen2011incremental, djafri2002spatio}.
\item[$\circ$] Allow partial or complete set of application data to store into tenant's private cloud while running the application in the public cloud \cite{balasubramanian2012security}.  It will create a few interesting research challenges in data cache, data optimization, data mining, and indexing of NoSQL data.  This allows flexible deployment scenario for data security and privacy.
\item[$\circ$] Allow tight integration of Security Operation Office~\cite{valeur2004comprehensive, wen2010lightweight, wen2012cafs}, Real Time Traffic analyser~\cite{zhang2015robust}, and Differential Privacy~\cite{dwork2006calibrating} into the {\em Trinity Model} to provide more application specific security and privacy protection.
\end{list}

%\section{Threats to Validity}
%\label{sec:threatsToValidity}
%\input{threats}

\section{Conclusion}
\label{sec:conclusions}
In this paper,  we analyze major issues of current DBaaS providers' models. We also propose {\em Trinity Model} to address these research challenges.  We discuss our ongoing work on conducting survey and interviewing with commercial cloud developers in order to have a thorough understanding of DBaaS providers and further substantiate our findings from the academic literatures in this paper.  We also discuss our future work on creating large scale SaaS services for retail industry using {\em Trinity Model} and research challenges in it.

Our work is, by far, the first kind to investigate DBaaS issues and propose an alternative more practical and suitable new DBaaS model to be used by one of the largest retail industry software providers for real world evaluation and impact the software industry (particularly in the SaaS domain for retail industry~\cite{pham2017paas}).

%\end{document}  % This is where a 'short' article might terminate

%ACKNOWLEDGMENTS are optional
%\section{Acknowledgements}
%\input{acknowledgements}


\bibliographystyle{abbrvnat}
\bibliography{DBaaSIssues}  % sigproc.bib is the name of the Bibliography in this case

\end{document}